\documentclass[12pt,epsf] {article}
\usepackage{amsfonts}
\usepackage{epsfig}
\begin{document}

\newcommand{\BOX}{\hfill $\Box$}
\newcommand{\NAB}{\hfill $\nabla \nabla \nabla$}
\newcommand{\BYDEF}{\stackrel{\Delta}{=}}

\newcommand {\beq} {\begin{equation}}
\newcommand {\eeq} {\end{equation}}
\newcommand {\bear} {\begin{eqnarray}}
\newcommand {\eear} {\end{eqnarray}}
\newcommand {\bears} {\begin{eqnarray*}}
\newcommand {\eears} {\end{eqnarray*}}
\newcommand {\done} {\quad\vrule height4pt WIDTH4PT}
\newcommand {\phantomI} {\vrule height20pt WIDTH0PT}
\newcommand {\barr} {\begin{array}}
\newcommand {\earr} {\end{array}}
\newcommand {\N} {\rm I\!N}

\def\Ooo{{ \Omega}_0^0}
\def\Oooc{{ \Omega}_0^{0,c}}

\def\bmr{\, \in \mbox{\boldmath $R$}}
                \def\dvl{\|_{2}^{2}}
                \def\hi{H_{\infty}}
                \def\eq{equation}
                \def\eqq{eqnarray}
                \def\eqn{eqnarray*}
                \def\lab{\label}
                \def\rn{{\bf R}^{n}}
                 \def\rmm{{\bf R}^m}
                 \def\rk{{\bf R}^k}
                   \def\rnm{{\bf R}^{n\times m}}
                    \def\rnk{{\bf R}^{n\times k}}
                    \def\ep{\epsilon}
                     
\def\limi{\mathop{\underline{\rm lim}}}
\def\star{*}
\def\cL{{\cal L}}
\def\cP{{\cal P}}
\def\tL{\tilde L}
\def\ztla{z_{\tau_l}^1}
\def\ztlb{z_{\tau_l}^2}
\def\intt{\int_{\tau_l}^{\tau_{l+1}}}
\def\summ{\sum_{n=\lf{\tau_l /\epsilon} + 1 }^{
\lf{\tau_l /\epsilon} + K(\epsilon ) } }
\def\cZ{{\cal Z}}
\def\ue{u_\epsilon}
\def\uy{{\ue(y)}}
\def\Qz{{\bf Q_0}}
\def\Qe{{\bf Q_\epsilon }}
\def\lf#1{{\lfloor #1 \rfloor}}
\def\lfe{{\lfloor \epsilon^{-1} \rfloor}}
\def\lims{\mathop{\overline{\rm lim}}}
\def\dfn{ \stackrel{\rm def}{=} }
\def\sp{\newline\noindent}
\def\sqr#1#2{{\vcenter{\hrule height.#2pt
      \hbox{\vrule width.#2pt height#1pt \kern#1pt
        \vrule width.#2pt}
    \hrule height.2pt}}}
\def\square{\vrule height6pt width7pt depth1pt}
\def\endpf{\hfill\square\bigskip}
\def\R{\mathop{\rm I\kern -0.20em R}\nolimits}
\def\Prf{\noindent{\bf Proof: }}
\def\fA{{\bf A}}
\def\fH{{\bf H}}
\def\ty{\tilde y^\epsilon}
\def\tz{\tilde z^\epsilon}
\def\fX{{\bf X}}
\def\cF{{\cal F}}
\def\cS{{\cal S}}
\def\tu{\tilde u}
\def\ts{\tilde s}
\def\cM{{\cal M}}
\def\fY{{\bf Y}}
\def\bV{{\bf Y}}
\def\R{\mathop{\rm I\kern -0.20em R}\nolimits}
\def\G{{\bf G}(g)}       \def\F{{\cal F}}
\def\C{{\cal C}}         \def\D{{\cal D}}
\def\nn#1{\mathop{\left |\kern -0.10em \left | #1
        \right |\kern -0.10em \right |} \nolimits}
\def\pd#1#2{\frac{\partial #1}{\partial #2}}

\def\uez{u_{\ep}\left(\bar{z}(t)\right)}
\def\exu{E_x^{u_{\ep}(\bar{z})}}
\def\dee{\Delta(\ep)}
\def\idee{\frac{1}{\dee}}
\def\tle{{\lfloor{\tau_l}/{\epsilon}\rfloor}}
\def\tl1e{{\lfloor{\tau_{l+1}}/{\epsilon}\rfloor}}

\newtheorem{definition}{Definition}[section]
\newtheorem{algorithm}{Algorithm}[section]
\newtheorem{lemma}{Lemma}[section]
\newtheorem{corollary}{Corollary}[section]
\newtheorem{proposition}{Proposition}[section]
\newtheorem{theorem}{Theorem}[section] 
\newtheorem{remark}{Remark}[section]
\newtheorem{example}{Example}[section]
\newtheorem{assumption}{Assumption}[section]         
\newtheorem{conjecture}{Conjecture}[section]         
\newtheorem{postulate}{Postulate}[section]
\newtheorem{hypothesis}{Hypothesis}[section]

\makeatletter
\def\EquationsBySection{\def\theequation{\thesection.\arabic{equation}}%
\@addtoreset{equation}{section}}
\makeatother
\EquationsBySection            

\title{{\bf
Linking Microscopic and Macroscopic Models for Evolution: Markov Chain Network Training and Conservation Law Approximations
}}
\author{
{\large \hspace{0cm} \vspace*{0.5cm}
Roderick V. N. Melnik\thanks{Tel.: +1-519-884-1970 (3662), Fax:
  +1-519-884-9738, E-mail: rmelnik$@$wlu.ca}
}
\\
{\large \hspace{-0cm} 
Mathematical Modelling and Computational Sciences, }\\  
{\large \hspace{-0cm} 
Wilfrid Laurier University, 75 University Avenue West,}\\  
{\large \hspace{0cm} 
Waterloo, ON, Canada N2L 3C5 and MCI, SDU, Denmark, DK-6400} \\  
}
\date{}
\maketitle
\thispagestyle{empty}

{
\abstract{In this paper, a general framework for the analysis of  a  connection between the training of artificial neural networks via the dynamics of Markov chains and the approximation of conservation law equations is proposed. This framework allows us to demonstrate an intrinsic link between microscopic and macroscopic models for evolution via the concept of perturbed generalized dynamic systems. The main result is exemplified with a number of illustrative examples  where efficient numerical approximations  follow directly from network-based computational models, viewed here as Markov chain approximations. Finally, stability and consistency conditions of such computational models are discussed.}

}


\bigskip

\section{Introduction}

Many concepts and tools used in information theory,  control, theory of approximation, dynamical systems and artificial intelligence are  closely 
 linked. Among such tools are neural networks.  The success in using neural networks in many application areas is well documented in the literature. This includes, but not limited to,  aeronautics, 
pattern and speech recognition (in particular in the context of various classification problems) \cite{Bishop1995,Corinto2006}, optoelectronics, robotics and autonomous navigation, determining structure-property relationships in material science applications \cite {Zhang1998,Ulmer1998},  constructing dynamic observers \cite{Ahmed2000},  identification and control of complex biotechnological cycles, industrial processes, and nonlinear systems in general \cite{Zhang1997,Cabrera1999}, modelling complex dynamic systems and nonlinear phenomena such as hysteresis  \cite{Gawin2001}. In many cases, a key to this success is kept by various  associations of (computational) neural networks with 
{\em the atomistic approach} to molecular design aimed at determining some relationships between the properties of the structure (thermomechanical, electromagnetic etc) and the structure itself described at  the microscopic (mesoscopic, and eventually  macroscopic) level.  Another set of tools and  successful approaches to structure modelling  is {\em the macroscopic approach} essentially based on the conservation law equations. In this contribution 
 we developed a general framework for establishing a link between neural network concepts and numerical approximations of conservation laws.

One of the major objectives in neural network theory is the study of constructive approaches to the design of effective learning algorithms \cite{Dasgupta2005}. We observe that models for learning of neural networks and models for the evolution of dynamic systems  are closely connected. The  idea of using neural networks as components in dynamic systems is now well established in the literature 
(e.g., \cite{Narendra1990}). Models for the dynamic interactions between evolution and learning have been used by a number of authors who use evolutionary algorithms for the optimization of neural structures. It is often argued that artificial neural networks (ANN) and evolutionary algorithms (EA) can be efficiently combined together. Indeed, if ANN is interpreted as a model for 
biological nervous systems and the learning process, then EA can be interpreted  as a model for biological evolution itself and the process of adaptation of large scales \cite{Sendhoff1999}. In this case, ANN becomes a useful mathematical tool for linking deterministic and probabilistic approaches in the approximation theory. This idea was developed in  \cite{Melnik1997,Melnik1998} where the evolution was proposed to be modelled by a generalized equation with training/learning rules (algorithms) understood in a probabilistic sense. The problem of learning in neural network theory is  formulated in terms of the minimization of an error function which is a function of adaptive parameters (weights and biases). In developing training/learning algorithms for ANN 
 we have to deal with incomplete data, and  an initial step of this procedure can be associated with the initializing of the weights in the network. Assigning these weights  random values \cite{Costa1999},  leads us to a probabilistic framework \cite{Wahba2000}.  Postprocessing of the available information is required for deriving deterministic models. For example, given a (training) set of input-output pairs $(u(i), y_i)$ ($i=1,...,n$ for n training/example subjects), we can ``smooth'' the data. The task is to construct a map which provides a good generalization (i.e. for given $u$ that does not belong to this map should provide a reasonable estimate/prediction of the unobservable output). This is reducible 
to finding  the best (functional) approximation to multivariate empirical 
(possible noisy, sparse, with some unobservable regions) data. However, if some  additional ``smoothing'' constraints are used,  we solve  not the original, but a regularized problem. This provides a link between standard (Tichonov-like) approaches to deterministic, but ill-posed, problems and statistical (e.g. the Bayesian) approaches \cite{Catala2000}. In this contribution we explore further this link between deterministic and probabilistic approaches in neural network theory by considering  learning and evolution in an intrinsic  dynamic connection.

The starting point of our discussion is a formalization of the learning/training process  in terms of the minimization of an error function based on the neural network approach.  In particular, starting with a minimal structure (no hidden layers), according to certain rules,   new connections, neurons, and layers are added.  Most commonly used are related to a (probabilistic)  adaptation of the network weights, a procedure which is intrinsically coupled with the overall network size (number of neurons on each layer and the total number of layers). Since  the size and topology of neural networks are closely connected with the required training time, in dealing with the training/learning problem 
we will have to  construct  {\sl dynamic} models. Such models arise naturally 
 in applications of ANN to optimal control problems in particular in those areas where system parameters are time varying \cite{Liao1999}. The approach developed in this contribution has common features with the on-loan evolution approach (previously discussed in the literature in the context  of  game theory and  control problems \cite{Agogino2000}) in a sense that our procedure can be interpreted as an on-line evolution algorithm applied to a certain 
 time-dependent equation.  The type of this equation is motivated by the fact that network architectures can be established by approximating the dynamic programming equations \cite{Balakrishnan1996}. In this paper, we also use a dynamic equation that corresponds to a certain neural network architecture and is derived from 
the framework of generalized dynamic systems (GDS) developed in \cite{Melnik1997,Melnik1998}.

In what follows, although we focus our attention on  feed-forward-based networks complemented by  backpropagation-like learning numerical procedures, the underlying procedure is the same for other architectures of neural systems where  we have to define the  connections between layers, the parameters (e.g., initial weights) and some learning rules. We demonstrate that these  connections can be established by combining  ``forward evolution'' (using neural networks) and `` backward evolution'' (using Markov chains associated with the original process).  This combination ensures the  minimization of  uncertainty in the following sense.  By processing information in response to discrete and continuous inputs, 
 the ANN allows us to model dynamic systems.  Since without a correction mechanism  uncertainty may increase dramatically  over time, we use Markov chains to construct an appropriate mechanism  where the architecture of the ANN depends  on the  cone of macroscopic events  
\cite{Melnik1998}. Although our approach is completely different in principle from  the classical  Gelenbe approach (see, e.g. \cite{Aguilar2001}), we note that in both cases 
 one has to exploit the idea of neural-network random structures. Since the original time-dependent models are  discretized in our approach,  one can interpreted the resulting  scheme via interactions among neurons so it is possible to  calculate probabilities of activation of network neurons in a way similar as it is done in the  Gelenbe approach.  

Finally, we note that when applying ANNs to complex dynamic systems, one of the most important issues   is to stabilize the learning mechanism \cite{Efe2000} (e.g., one has to avoid the unbounded growth of the adjustable parameter in ANN). We show in this paper that 
 this  non-trivial task can be linked directly to discretizing  conservation law models, and that the ANN stability can be treated effectively through the stability of numerical approximations of conservation laws, rather than through the  Lyapunov stability (e.g. \cite{Levin19931996}).


\section{Systems Theoretic Framework for Constructing Neural Networks and Dynamic Training Rules}

By applying neural networks  to learning and identification of dynamic
 systems, we attempt to influence the behavior of these  systems by some components built-in into the systems  \cite{Narendra1990,Narendra1996}.  Hence, the procedures for controlling such dynamic systems using ANN  should be connected 
 in one way or another with approximations of 
system dynamics. Moreover, since control, in addition to the requirements of fast and accurate, should ensure stability and robustness of the system, the stability of control is closely connected in such cases with the stability of numerical approximations of models describing the systems dynamics.  

Recall that in most typical situations a nonlinear dynamic system can be described from a systems theoretic  point of view by the following set of equations
\begin{eqnarray}
\left \{
\begin{array}{l}
x(k+1) = f[x(k), u(k)], \quad f(0,0) =0, \\
y(k) =h[x(k)], \quad h(0)=0,
\end{array}
\right.
\label{2eq1}
\end{eqnarray}
where $u(k)$, $y(k)$ $\in \mathbb{R}^m$ and $x(k) \in \mathbb{R}^n$ are input, output, and state vectors, respectively, at discrete time $k$, and  mappings $f: \mathbb{R}^n \times \mathbb{R}^m \rightarrow \mathbb{R}^n$ and $h: \mathbb{R}^n \rightarrow \mathbb{R}^p$ are given functions, while  control based on the state vector and control based only on input-output data needs not to be the same \cite{Narendra1996}. Model (\ref{2eq1}) provide adaptive means for implementing digital neuro-controlled systems and for  further insight into the neural learning and adaptation. We note that instead of  model (\ref{2eq1}),  the dynamic system model can be determined through identification (e.g., with multi-layered neural networks) using the data obtained by a recurrence equation describing the output signal and taking into account the internal perturbations (which influence the output signal):
\begin{eqnarray}
y(k+1)= \tilde{h}(x(k+1), y(k), u(k), e(k+1)).
\label{2eq2}
\end{eqnarray}
New output $y(k+1)$, determined from the state vector $x(k+1)$ by using the given input data  $[y(k), u(k)]$, will introduce error $e(k+1)$,
 representing internal perturbations at  time $k+1$. For example, it can mimic the effects of synaptic time delay representing the ``memory'' of neurons which may differ from connection to connection.  Mapping $\tilde{h}$ in (\ref{2eq2}) can be viewed as  a perturbed function $h$. In principle, this equation, which combines the information given by model (\ref{2eq1}) is sufficient to describe the training procedure. From a numerical point of view such a nonlinear-control-framework 
consideration  \cite{Narendra1996}  can be viewed as an approximation to the first equation in (\ref{2eq1}), describing the evolution of states of the system, supplemented by an approximation of the ``constitutive'' law (the second equation in (\ref{2eq1})), describing some properties of the system, control rules, and/or requirements imposed by the user. Hence,  if the original description of the system is set by using control-theoretic approach, model (\ref{2eq2}) 
 can be viewed as an approximation to the Hamilton-Jacobi-Bellman-type (HJB) equation describing some ``energetic'' characteristic of the system such as cost or value function \cite{Melnik1997,Melnik1998,Melnik2003}. If the entire controlled process is modelled by the ANN, then the  training/learning process for  the neural network (using specific outputs with each of several inputs) can be
 associated  with the dynamics of a certain conservation law model that describes the evolution of the nonlinear system. In this case,  the inputs of ANN are associated  with the initial conditions of the conservation law model. 
The description of the output-input training process  will be  understood here as a constructive backpropagation based on a process-associated Markov chain approximation. This approximation will be linked to numerical approximations of conservation laws.

Following  \cite{Narendra1996},  we consider  a neural network as a conveniently parameterized class of nonlinear maps. In the next paragraphs we develop the methodology for linking this class of maps  to  conservation law approximations. In this context it is important to note that when  a network is chosen to approximate a given mapping using IO data, the IO set is finite, whereas  a conservation law defines an infinite set of input-output data. However, once such a conservation law model is approximated the process of neural network training and the approximate conservation laws can be described within the same general framework based on the concept of generalized dynamic systems \cite{Melnik1997,Melnik1998}. In what follows,  our  consideration will be pertinent to discrete-time (even though the original system might be continuous in time), which reflects the fact that most of complex systems are controlled by computers and therefore it is quite natural to consider approximations to complex dynamic systems as being discrete in time \cite{Narendra2000}. Moreover, any model, no matter how sophisticated it is, should account for uncertainties of the environment \cite{Melnik1997,Melnik1998}  which are easier to deal with in the discrete time formulations \cite{Narendra2000}. Furthermore, the stability issues are much more transparent in the discrete space-time of events where dynamic systems evolve. Finally,  in the context of neural networks the discrete-time consideration is the most natural way to proceed with the analysis because  neural networks are proved to be universal approximators and any continuous function can be approximated arbitrarily well on a compact set with a (multilayer feedforward, radial basis function, or another) neural network (e.g., \cite{Cybenko1989,Hornik1989,Barron1993,Cabrera1999}).


\section{Approximating System Hamiltonians with Neural Networks}

If the dynamic system under consideration is Hamiltonian, the problem of training neural networks with dynamic system rules governing the  evolution of this system requires efficient procedures for approximating the Hamiltonian of such a system by a neural network. As a starting point, let $T$ be a given set of times during which the network is trained, $\Sigma$ is a state space of the dynamic system, $U_T$ is a set of all permissible strategies for training, and $X_T$ is the domain of definition of the system Hamiltonian assumed to be a compact Borel set \cite{Croft1991}. Further constructions are based on the following fundamental property of
applications of neural networks in the theory of approximation \cite{Cybenko1989,Hornik1989,Barron1993}:

\begin{theorem}
{\sl If $H \in \mathbb{L}^1(X_T)$, 
then for any arbitrary small 
$\epsilon>0$  there exists a network $\tilde{H}$ such that
\begin{equation}
||H- \tilde{H}||_{\mathbb{L}^1(X_T)}< \epsilon.
\label{3eq1}
\end{equation}
}
\end{theorem}

For the description and control of complex dynamical systems via neural networks both feedforward (FNN) evolutionary networks (for the modelling of nonlinear input-output dynamics) and feedback evolutionary networks (for the implementation of training algorithms) are required.  The link between those can be established on the basis of the Markov chain approximations as follows. 
First, recall that  if $H$ is a continuous function we can construct a uniform approximation (e.g., by using a FNN \cite{Cybenko1989,Hornik1989}). Note also that if smoothness of the function $H$ is measured in terms of its 
Fourier representation an actual estimation of the network
performance can also be obtained (e.g., \cite{Barron1993}). Therefore, it is natural to associate a FNN model with the approximation  $\tilde{H}$ of the system Hamiltonian (for simplicity,  with one layer of sigmoidal nodes or units).  In $\mathbb{R}^n$ this model can 
be implemented by the following functions  
\begin{equation}
 H_n({\bf x})= \sum_{i=1}^n \alpha_i x_t({\bf y} \cdot {\bf x} + \beta_i)
+ \alpha_0 
\label{3eq2}
\end{equation}
with ${\bf x}, {\bf y} \in \mathbb{R}^n, \; \alpha_i, \beta_i \in  \mathbb{R}.$
The point that we are making here is that such models alone   may be of limited applicability when applied in 
uncertain dynamic  environments, because the quality of network performance  
is  dependent strongly on its  training capabilities with respect to the regularity of $H$ and the structure of $X_T.$  In considering computational models of system dynamics the structure of the network-approximator like (\ref{3eq2}) cannot be considered to be fixed, but rather it has to be adapted to  the change of environment and internal perturbations. As a consequence, the most appropriate approximation  in the framework of (\ref{3eq1}) can lead to a situation where activation function $x_t$ may not be continuous,
subject to regularity of $H$. Furthermore,  even if this function  is continuous, in order to account for the topological structure of $X_T$ each network layer has to be characterized by both feedforward and feedback operators. 

\begin{remark}
In a special case, where $X_T \equiv \mathbb{R}^N$ and hence $H:   \mathbb{R}^N \rightarrow \mathbb{R}^M$ ($N, M \in \mathbb{N}$),
such operators may be interpreted through  weight matrices 
\cite{Santini1995}. 
\end{remark}

The situation we described above is of  the same nature as in 
control theory where a regularity balance between control and value 
functions has to be achieved \cite{Melnik1997A,Melnik1997}. In the context of our framework, this balance is achieved by simultaneous computational treatments of the ``feedforward'' and ``backward'' evolutionary processes, which can be effectively implemented  by using along with feedforward approximators feedback networks such as block feedback networks (BFN). Below we explain this idea in detail. 

Consider  Hamiltonian $H$ mapping $X_T$ into $\mathbb{R}^M$ and its ``computational equivalent'', that is a Turing computable function $H_T$ such that
\begin{equation}
 H: X_T \rightarrow \mathbb{R}^M, \; \mbox{and} \; 
H_T:  \mathbb{N} \rightarrow \mathbb{N}.
\label{3eq3}
\end{equation}
Then, since BFN models have the same computing power as Turing machines (e.g., 
\cite{Santini1995}), we conclude that  there exists a BFN ${\tilde{H}}_T$ 
such that for any input $n$ from a subset of  $\mathbb{N}$
a finite number of network steps produces $H_T(n)$. 
Therefore, due to the G\"{o}del numeration procedure (e.g., \cite{Chaitin1982} and references therein),
  any Hamiltonian function
 $H: \mathbb{R}^N 
\rightarrow \mathbb{R}^M$ can be arbitrary well approximated by a 
network implementation of  function $H_T$. Since both feedforward $\tilde{H}$ and feedback $\tilde{H}_T$ networks provide an approximation to $H$, the development of constructive algorithms for training requires further study of the connection between   $\tilde{H}$ and $\tilde{H}_T$. This connection between the concepts pertinent to feedforward and feedback networks is often overlooked. Examples where this connection  becomes  transparent are provided by associative memory networks or recurrent neural networks, where feedforward connections should be supplemented by some ``memory'' neurons or synaptic time delay, which may vary from connection to connection. A natural way to establish such a connection is to construct a Markov chain associated with the system evolution in such a way that it reflects  the process of network training on the problem specific information.

Now, we are in a position to formulate the problem of training  in terms of approximating the system Hamiltonian in such a way that the necessity of applying both forward and backward dynamic  rules is transparent. For this purpose we consider a relatively simple case, where we assume that 
\begin{equation}
x_t: \mathbb{R} \rightarrow \Sigma
\label{3eq4}
\end{equation}
is a sigmoidal function, meaning that
\begin{equation}
\lim_{t \rightarrow -\infty} x_t=0, \; 
\lim_{t \rightarrow +\infty} x_t =1.
\label{3eq5}
\end{equation}
The function $x_t$ here is the activation function for a neural network  defined by its neurons as the following mapping
\begin{equation}
 x_t \circ \mu: T \otimes \Sigma \otimes U_T \rightarrow \Sigma, 
 \label{3eq6}
\end{equation}
where $\mu: T \otimes \Sigma \otimes U_T \rightarrow \mathbb{R}$ is known as the  decision maker (DM) function \cite{Melnik1998} such that it has to be adjusted (trained) so that the neural network (\ref{3eq6}) leads to an approximation of $H$.
If the network (\ref{3eq6}) is trained with some dynamic rules by using a dynamic system whose Hamiltonian is $H$, then the process of approximation of function $H$ in terms of (\ref{3eq1}) can be seen as the construction of a training strategy for a new network  $\tilde{H}_n^{\epsilon}$ depending on the arbitrarily small positive parameter $\epsilon$ and arbitrarily  large number of sigmoidal nodes of the associated network $n$ (e.g., (\ref{3eq2})) so that 
\begin{equation}
||H- \tilde{H}_n^{\epsilon}||_{\mathbb{L}^1(X_T)} \rightarrow \min,
\label{3eq7}
\end{equation}
where  $H \in \mathbb{L}^1(X_T)$. The main difficulty in the solution of problem (\ref{3eq7}) stems from 
 a priori unknown character of dependency of the network on parameters which determine the 
function $\mu$.  In the most general setting, the problem of constructing the mapping $\mu$ is intrinsically 
connected with the definition of dynamic rules 
in singular stochastic control problems \cite{Melnik1997}, and in interpreting (\ref{3eq7}) one expects that  
\begin{eqnarray}
 \mbox{if} \; \; 
\epsilon \rightarrow 0^+ \; \; \mbox{and} \; \; n \rightarrow \infty \; \;
\mbox{then} \; \; \tilde{H}_n^{\epsilon} \rightarrow H. 
\label{3eq8}
\end{eqnarray}
However, for the constructive solution of (\ref{3eq7}) by using neural networks we need additional information. For example, if FNNs are applied to the solution of the problem, we need information  on a coupling rule between $\epsilon$, $n$, and the topology specified 
 by $X_T$.  This coupling rule will determine conditions for  the system stability. 
 If BFNs are applied  to the approximation of $H$ on an arbitrary set 
$X_T$, we need  additional information on the network dimension and architecture \cite{Santini1995}. This information can be made available only in {\em a sequential manner}, and the appropriate tool for the analysis of the network performance in such situations is  a family of Discrete Markovian Decision Processes (DMDP). In this case  a  model for
the training process, written in terms of the decision maker function $\mu$,  depends on Markov Chain 
parameters with possible 
 discontinuities that are dependent on values of the sigmoidal 
function $x_t.$  If the functional dependency of $\mu$ is chosen a priori, this may lead  to the underestimation of 
possible irregularities of the function $H$ which can happen if, e.g., smoothness of $H$ is measured in terms of its Fourier representation \cite{Barron1993,Muller1993}. In the next section we aim at constructing a  model that gives an approximation to $\mu$ with respect to some learning 
rules determined from the Markovian property of the process
$(x_t, \mu)$.

\section{Modelling Dynamics of Network Training}

From a constructive approximation point of view, the training process of neural networks describing dynamic systems evolution can be seen as the solution to the following problem. We have to construct a model (possibly, a hierarchy of models \cite{Melnik1997A}) for 
the decision maker function $\mu$ 
in such a way that a neural 
network approximates the Hamiltonian of a dynamic system evolution, while the dynamics of this evolution 
is described by  the process of network training.

\begin{center}
\begin{figure}
\centerline
{\epsfig{file=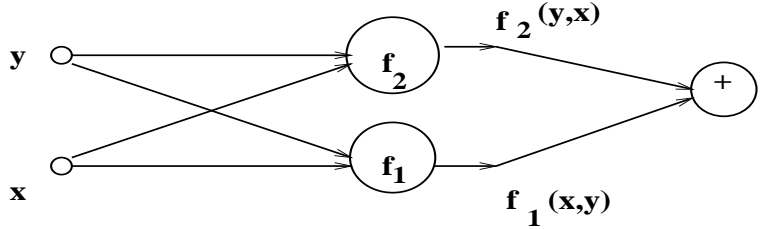, height=3cm, width=10cm}} 
\caption{A typical unit of network architecture}
\label{fig1}
\end{figure}
\end{center}

Any two sets of input-output data in the training process can be represented by  the mathematical model of a dynamic system that couples two space-time events of the system evolution, $e_n$ and  
$e_{n+1}$, by a function of the perturbed velocity $v_{\epsilon}$ and the system Hamiltonian or its approximation $H$: 
\begin{equation}
 e_{n+1}=H(v_{\epsilon}, e_n), \; \; n=0,1,... 
 \label{4eq1}
 \end{equation}
 The perturbed velocity is introduced to account for the changing environment and/or internal perturbations \cite{Melnik1998}. Then, if we specify  a sequence of events $(e_0, e_1,...)$ by temporal evolution and formalize the dynamics of the system by a discrete-time model, we obtain the following two equations  \cite{Melnik1998}
\begin{equation}
 \left \{ \begin{array}{l}
x_{t+1}=H_1(v_1,x_t), \\   
        h_{\tau+1}=H_0(v_0,h_{\tau}), 
        \end{array} \right.   
\label{4eq2}
\end{equation}
where $H_1$ is an approximation to $H$ and $\displaystyle v_1 = \lim_{\varepsilon \rightarrow 0^+} v_{\varepsilon}$,  $H_0$ is an operator for sequential corrections of such an approximation needed due to system external/internal perturbations and $v_0$ is the function characterizing the rate of such perturbations. The dynamics of the system described by (\ref{4eq2}) operates on two coupled temporal scales, $t$ and $\tau$ with spatial trajectories described by 
$x_t$ and $h_{\tau}$, respectively.
Since perturbations (internal and/or external) is an intrinsic part of any dynamic system, we associate the class of neural networks with built-in training capability (determined by models for $\mu$)
as Infinite Length Perturbed Markov Chains (ILPMC)  \cite{Melnik1998}. Since the form of the functional dependency of $\mu$ cannot  be a priori fixed, the weights and the network structure  may change  which leads to a situation much more complicated 
 than traditionally dealt with (see Fig. \ref{fig1}). Some algorithms dealing with such a situation are  known (e.g., cascade-correlation, pruning algorithms, etc). However, our approach is different from the previous developed since it is based on 
 the Markovian character of the process $(x_t, \mu)$, associated with both system dynamics and the process of training, and the concept of  generalized dynamic systems (with appropriate  assumptions on dynamic stochastic rules \cite{Melnik1998}). In our approach, by using conditional probabilities of the Markov chain, the velocity function between two macroscopic events in the evolution of the generalized dynamic system is introduced as a measure of changes which take place on the microscopic level with respect to the macroscopic behavior of the system. This velocity function provides a link between microscopic and macroscopic models for the evolution of dynamic systems.
Similar to model (3.7), models (\ref{4eq1}) and (\ref{4eq2}) 
 are interpreted in terms of  the  limit   $\epsilon \rightarrow 0^+$
and $n \rightarrow \infty$, and the connection between them can be analyzed when  learning rules for the system evolution in response to perturbations are  introduced. In particular, in the limit (\ref{3eq8}) both sequences $(x_0, x_1,...)$ and $(h_0, h_1,...)$ merge.  As an important part of the system dynamics, perturbations  can be formalized   from the very beginning of the modelling process
 as  a decision making process with limited information available.  

\begin{remark}
Since perturbed and unperturbed models might give rise to  qualitatively distinct types of descriptions of system behavior for any arbitrary $\epsilon >0$,  in  \cite{Melnik1998} it was already emphasized that the perturbation parameter alone cannot be an appropriate characteristic of the model's uncertainty, and  this fact results in two informational sequences in model (\ref{4eq2}) considered on {\em two different time scales}. 
\end{remark} 

The learning rules are introduced into the model by 
the equation written in terms of function $\mu$  (Section 3). 
In particular, following \cite{Melnik1998,Melnik2003}, we arrive at 
\begin{proposition}
If   $H \in \mathbb{L}^1(X_T)$, then the process of training/learning can be  associated with
the following equation:
\begin{equation}
(1+v_1) \left [ \frac{\partial \mu}{\partial x} + \frac{1}{v_1}
\left ( \frac{\partial \mu}{\partial t} + f_0 \right ) \right ] =0,
\label{4eq3}
\end{equation}
where, in the context of neural networks, $v_1$ is the velocity of information transmission  
between neurons, and $f_0$ is a training/learning goal defined by a priori knowledge/assumptions
on $X_T$ and the function $H$. 
\end{proposition}

This process can be approximated by  an appropriately constructed Markov chain associated with the evolution of the dynamic system under consideration. Since  function $\mu$ is required 
to be adapted to  a priori given information, the practical implementation of training procedures may lead to approximations of the network-activator functions  by a piecewise-deterministic stochastic process. Such non-diffusion stochastic models have been previously studied in theory of DMDP and in theoretical physics (see \cite{Melnik1998} and references therein). Such models are multiscale models for the evolution of dynamic systems described by (\ref{4eq2}) and, as pointed out in \cite{Melnik1998}, can be  interpreted by a system of coupled differential equations on two scales:
\begin{eqnarray}
\left \{
\begin{array}{ll}
     \dot{h}(\tau)=v_0(\tau, h, \mu), &  \\
\displaystyle
    \frac{\partial \mu}{\partial   t}+v_1(t, x, \mu)   
 \frac{\partial \mu}{\partial x} =0.  &
 \end{array} 
 \right.
 \label{4eq4}
 \end{eqnarray}
We conclude this section with the following observation.
\begin{remark}
Both parts of the perturbed velocity  functions $v_0$ and $v_1$
inherit their dependency on the decision-maker function $\mu$.
If two events between which GDS evolution
has to be studied   are specified (e.g., two sets of input-output data are entered), then   a pair of functions
     $(h(\tau), \mu(t,x))$  gives the solution to the training problem under consideration.
\end{remark}
Next, we describe a procedure that allows us to approximate this pair.

\section{Network-Based  Computational Models for Dynamics as Markov Chain Approximations.}

Although the perturbed system dynamics $x_t^{\epsilon}$ might not be governed by the Markovian property (and the function $x_t^{\epsilon}$ might not be sigmoidal), the pair of functions $(h(\tau), \mu(t,x))$ does possess the Markovian property \cite{Melnik1998}. Therefore, the key idea here is based on the construction of a Markov chain approximation  simultaneously with an approximation of the system dynamics (which depends on Markov chain parameters). This allows us to guarantee system  stability and to derive stability conditions in explicit form. The Markov chain will play in our constructions the role 
of a ``training/learning'' rule for the system dynamics considered in the case where the perturbed system's velocity is replaced by its  approximation (function $v_1$) in the macroscopic (or decision maker's) frame of reference.

More precisely, we will approximate the pair of functions $(h(\tau), \mu(t,x))$ (which describes the process of GDS evolution and possesses the Markovian property)  by a pair of discrete functions
\begin{eqnarray}
(h(\tau), \mu(t,x)) \rightarrow (\xi_n^{\tau
h}, {\mu}_n^{\tau h}),
\label{5eq1}
\end{eqnarray}
where $\xi_n^{\tau h}$
is an   associated  (with the  microscopic   frame  of  reference)
Markov Chain state.

First, we consider model (\ref{4eq1}) describing the evolution of dynamic systems in discrete space-time of events. Let an elementary space-time cell be 
$c_{ij}$ and the complete spatio-temporal region, where the systems dynamics is studied, be $\bar{G}$ such that two events 
\begin{eqnarray}
e_j, \; e_{j+1} \in 
 c_{ij} \equiv [x_i,x_{i+1}] \otimes [t^j,t^{j+1}] \subset \bar{G} 
\label{5eq2}
\end{eqnarray}
of    system
evolution are governed by the process $(x_t, \mu_t)$. Then
 we specify these events by
two
pairs of    discrete    functions    
\begin{eqnarray}
e_j=(\xi_j^{\tau     h},
\mu(x_i,t^j)), \; \; e_{j+1}=(\xi_{j+1}^{\tau     h},
\mu(x_{i+1},t^{j+1})),
\label{5eq3}
\end{eqnarray}
 where $\xi_j^{\tau h}=x_i^j$ and
$\xi_{j+1}^{\tau h}=x_{i+1}^{j+1}$  are  states  of  the  associated
Markov Chain.

Next, we consider  the perturbed generalized dynamic system whose training/learning   dynamics can be described by the following equation, obtained from (\ref{4eq3}) or (\ref{4eq4}) under appropriate assumptions
\begin{equation}
     \frac{\partial    \mu}{\partial     t}     +
v_1(t,x, \mu) \frac{\partial \mu}{\partial x} =\tilde{f}_0(t,x, \mu)
\label{5eq4}
\end{equation}
with the  approximation of the initial condition 
  in the DM-time scale 
\begin{equation}
\mu(x,t)|_{t=t_0}=\delta(\epsilon), 
\label{5eq5}
\end{equation}
where $\delta(\epsilon)$ is a set of initial conditions with
$\epsilon$ dependent on the approximation of the function 
$v_0$ in the coupled system (\ref{4eq2}) (or (\ref{4eq4})).

To preserve basic macroscopic  features  of
the system, the values of jumps $\Delta \xi_j^{\tau h}=
  \xi_{j+1}^{\tau  h}  - \xi_j^{\tau h}$ of this
chain should be subordinated to the corresponding
approximation of  system-environment  boundaries.
This can be done via establishing the connection between the GDS and perturbed GDS.  
\begin{definition}
Let (\ref{5eq2}),  (\ref{5eq3})
  be  two  subsequent
macroscopic events of GDS evolution that are taking place
with probability 1.
Then the    GDS  velocity function between the macroscopic
events $e_j$ and $e_{j+1}$
can be defined in an elementary space-time cell $c_{ij} \subset \bar{G}$
as 
\begin{equation}
 v(t,x)=  \lim_{\tau  \rightarrow  0}  \frac {E^{\tau
h|(x_i,\mu^j)} \Delta \xi_j^{\tau h}} {\tau},
\label{5eq6}
\end{equation}
 where    the numerator under the limit in (\ref{5eq6}) is the velocity of
the Markov  Chain ($v_{\rm MC}$) between two  subsequent  macroscopic  events.
\end{definition}

We observe that 
\begin{remark}
If  $\displaystyle  \lim_{\epsilon \rightarrow 0^+} v_{\epsilon}=v_1$ then
     \begin{equation}
    \lim_ {   {\epsilon  \rightarrow  0^+,}  \;  {n  \rightarrow
\infty} }
v(t,x) = v_1,
\label{5eq7}
\end{equation}
and together with (\ref{4eq1}) this defines   an  Infinite  Length  Unperturbed  Markov  Chain
(ILUMC). 
\end{remark}
Since perturbations cannot be ignored, (\ref{5eq7}) can be fulfilled only approximately, which leads to the consideration of 
an ILPMC, the concept already mentioned in Section 4.

Before formulating consistency conditions for the Markov chain associated with the dynamics of training rule (\ref{5eq4}), (\ref{5eq5}) we recall \cite{Melnik1998} that 
\begin{definition}
The state space in the initial moment of observation in the macroscopic frame of reference is defined as
\begin{eqnarray}
\Xi(i;0) = \{x_i, i=0,1,2,...,N; \quad N=2n, \quad n= \lceil (T-t_0)/h \rceil \},
\label{5eq8}
\end{eqnarray}
and subsequently 
the  cone of macroscopic events of system evolution
is defined by a    set   of   macroscopic   events  as  
   a mapping from the minimal resolution set for the identification of all macroscopic events relevant to the system evolution in the limit $n \rightarrow \infty$  to   $\rightarrow \Xi(i;j),$
     where
\begin{eqnarray}
     \Xi(i;j) =  \{(x_i,t_j),  \quad i=\overline{k,2n-k},
\; 
j=k,\; k=\overline{0,n} \}.
\label{5eq9}
\end{eqnarray}
\end{definition} 

Now we are in a position to formulate 
\begin{proposition}
\underline{\em The consistency conditions} (local and global, respectively) of  the  
  Markov Chain $\xi_n^{\tau h},  \;
n< \infty$    with   the    Markov    process
$(h(\tau),\mu(t,x))$, defined by the mathematical model of GDS evolution
(\ref{5eq4}), (\ref{5eq5}), are:
\begin{eqnarray}
& &
 E_{n}^{\tau h |(x_i,\mu^j)} \Delta \xi_j^{\tau h} =
     v_1(x_i, t^j, \mu^j) \tau + o(h+ \tau)
     \label{5eq10}
\\[10pt]
& &
    cov_{n}^{\tau h |(x_i, \mu^j)} \Delta \xi_j^{\tau h} =
      o(h+ \tau).
      \label{5eq11}
\end{eqnarray}
\end{proposition}

Using these ideas and following  \cite{Melnik1998}, a simple approximation to 
(\ref{5eq4}), (\ref{5eq5}) can be constructed and in the next sections we generalize this approximation to a large class of schemes for approximating the dynamics of training rules in neural network applications. Let us first highlight 
the main steps for the construction of  the scheme proposed in \cite{Melnik1998}. 
\begin{itemize}
\item
In the cone of macroscopic events we introduced a
floating grid:
\begin{eqnarray}
\omega_{\tau h}^{\triangle}
      =  \{(x_i,t_j^{\tau_{j-1}} ),  \ i=\overline{k,2n-k},
\; j=k, \; k=\overline{0,n} \},
\label{5eq12}
\end{eqnarray}
     where $ t_j^{\tau_{j-1}}=t^{j-1}+\tau_{j-1}$ when $j>1,$
$t_j^{\tau_{j-1}}=t^0+\tau$ when $j=1,$ and
$t_j^{\tau_{j-1}}=t^0$ when $j=0$,
where  we constructed 
 the following  discrete  scheme (based on upwind approximations with flux limiters \cite{Quarteroni2000}) 
\begin{eqnarray}
     d_i^{j+1} & = & d_i^j \{1-\frac{\tau}{h}[|v|+v^-   \gamma_4-v^+
\gamma_1] \}
     +
 \frac{\tau}{h}d_{i-1}^j \{[v^+(1+\gamma_2)+v^-\gamma_4] \}  
     +
\nonumber \\
& &
     \frac{\tau}{h}d_{i+1}^j \{[v^-(1-\gamma_3)-v^+\gamma_1] \}
     +
\frac{\tau}{h}d_{i-2}^j \{[-v^+\gamma_2] \}
     +\frac{\tau}{h}d_{i+2}^j \{ v^- \gamma_3 \},
     \label{5eq13}
\end{eqnarray}
where $d$ is a discrete function approximating function $\mu$ on grid (\ref{5eq12}).
\item
It is easy to see that the sum of all coefficients near the unknown function on the right-hand side (\ref{5eq13}) gives the unity. This fact allows us  associate these coefficients with transition probabilities of a Markov Chain,
 provided those coefficients are nonnegative:
\begin{eqnarray}
\left [
\begin{array}{l}
     1-  \frac{\tau}{h}(|v|+v^-  \gamma_4  -  v^+ \gamma_1)
\geq 0, 
\quad \gamma_2 \leq 0, \quad \gamma_3 \geq 0,
\\[10pt]
 v^+(1+ \gamma_2)+v^- \gamma_4 \geq 0, \quad
      v^-( \gamma_3-1)+v^+ \gamma_1 \leq 0.
\end{array}
\right.
      \label{5eq14}
\end{eqnarray}
\item
 The consistency conditions  (local and global, respectively) of the Markov Chain  (defined by time-transitions of the  discrete
scheme (\ref{5eq13}))  with the process $(h(\tau),  \mu(t,x))$ (defined by
the model (\ref{5eq4}),(\ref{5eq5})) couple flux limiters of the scheme
\begin{eqnarray}
&&
     \tau[ v^-(1-\gamma_4+\gamma_3)            -v^+(1+\gamma_1
-\gamma_2)-v]= {\it o} (\tau+h),
\label{5eq15}
\\
&&
\tau \{h [v^+(1-\gamma_1-3 \gamma_2) +v^-(1+\gamma_4+3\gamma_3)]
-\tau v_{MC}^2 \}= {\it o} (\tau+h),
\label{5eq16}
     \end{eqnarray}
where the Markov chain velocity in this case is determined as $v_{MC}= v^{-}
(1 - \gamma_4 + \gamma_3) - v^{+} (1 + \gamma_1 - \gamma_2)$, and ${\it o}$ is the Landau symbol. Using the idea of probabilistic characteristics 
the term ${\it o} (\tau+h)$ in (\ref{5eq15}) and (\ref{5eq16}) can be eliminated.  The  nonnegativeness  of   covariance leads to an additional stability condition 
\begin{equation}
     \frac{\tau}{h} \leq \frac{v^+(1-\gamma_1-3 \gamma_2)+
     v^-(1+\gamma_4+3 \gamma_3)}
     {{[v^-(1+\gamma_3-\gamma_4)-v^+(1+\gamma_1-\gamma_2)]}^2},
     \label{5eq17}
\end{equation}
which can be satisfied under appropriate choice of the flux limiters. As a result, we have arrived to the stable Markov chain approximation of the training process. 
\item
More precisely, if the interpolation interval $\tau$ is such that  conditions (\ref{5eq14}) and (\ref{5eq17}) are satisfied and the transition probabilities of the Markov Chain $(\xi_n^{\tau h},  \; n <
\infty)$ are taken in the form
\begin{eqnarray}
       p^{\tau   h}[x_k^j,
x_i^{j+1}|d(x_k^j,t^j)] =
     \left\{ \begin{array}{ll}
     1-\frac{\tau}{h}[|v|+v^- \gamma_4 -v^+\gamma_1],  & k=i,  \\
     \frac{\tau}{h}[v^+(1+\gamma_2)+v^-\gamma_4],  & k=i-1,  \\
     \frac{\tau}{h}[v^-(1-\gamma_3)-v^+\gamma_1],  & k=i+1,  \\
    - \frac{\tau}{h}(v^+\gamma_2),  & k=i-2,  \\
     \frac{\tau}{h}(v^-\gamma_3),  & k=i+2,     \\
     0, &    \mbox{    otherwise},
     \end{array} \right.
\label{5eq18}
\end{eqnarray}
$\forall j=\overline{0,n-1}$ and $i=\overline{j,N-j}$
($\gamma_2=0  \; \mbox{for} \; i=j \; \mbox{and} \;   \gamma_3=0
\; \mbox{for} \; i=N-j $),
then 
     the Markov    Chain    approximation   of   the   process
$(h(\tau),\mu(x,t))$ is
stable, and 
\begin{eqnarray}
    d(x_i^{j+1},  t^{j+1})= 
    \sum_k p^{\tau   h}[x_k^j,
x_i^{j+1}|d(x_k^j,t^j)] d(x_k^j,t^j).
\label{5eq19}
\end{eqnarray}
\end{itemize}

A numerical procedure like (\ref{5eq19}) is an explicit (evolution forward)
stabilization procedure  where  the  DM-function is a stabilizing
factor subject to the velocity of the system. Moreover, 
 when $n \rightarrow \infty$ the velocity
of  the  Markov
Chain converges  to the velocity of the process in the sense of
the Markov theorem (e.g., \cite{Melnik1998}). This puts the approach used here on a rigorous theoretical basis. In the next section we show that this approach can be generalized to an important class of conservation-law-based models for the training process of neural networks.

\section{Network-Based Probabilistic Approach to Conservation Law Approximations}

The network-based methodology  developed  in the previous sections can be  applied to conservation law approximations. In what follows we demonstrate this on a number of practical examples where we show that many efficient finite difference schemes for the approximation of conservation law equations follow directly from our technique. 

Consider a model analogous to that of training/learning dynamics for perturbed GDS, but with $\mu \rightarrow u$, $v_1 \rightarrow a$, and $\tilde{f}_0 \equiv 0$:
 \begin{eqnarray}
\frac{\partial u}{\partial t} + a(t, x, u) \frac{\partial u}{\partial x} =0,
\label{6eq1}
\end{eqnarray}
where $u$ is some physical quantity of interest, and $a$ is the velocity related to the transport of that quantity. With the Cauchy initial condition 
\begin{eqnarray}
u(x, 0) = u_0(x), \quad x \in \mathbb{R}
\label{6eq2}
\end{eqnarray}
this equation gives a model of  a physical system without dissipation in the form easily amenable to a nonlinear (quasi-linear) conservation law. Indeed if $a(u)= F'(u)$, then model (\ref{6eq1}), (\ref{6eq2}) is equivalent to
\begin{eqnarray}
\frac{\partial u}{\partial t} + \frac{\partial F}{\partial x} =0, \quad  u(x,0)= u_0(x), \quad x \in \mathbb{R}.
\label{6eq3}
\end{eqnarray}
This model and model (\ref{5eq4}), (\ref{5eq5}) describing the dynamics of network training belong to the same class of mathematical models, and such models provide building blocks for many mathematics applications  in science and engineering. One of the sources of difficulties for numerical solutions of such problems is that 
 in many practical situations the solution to (\ref{6eq3}) might not be continuous, and one has to deal with possible discontinuities across certain curve $x= \sigma(t)$. In other words, if
\begin{eqnarray}
\lim_{x \rightarrow \sigma(t)^{-}} u(x,t) =u_{\rm left}, \quad 
\lim_{x \rightarrow \sigma(t)^{+}} u(x,t) =u_{\rm right} \quad \mbox{and} \quad  u_{\rm left} \neq u_{\rm right},
\label{6eq4}
\end{eqnarray}
one should be able to select a physical meaningful solution from a set of all possible solutions (since two or more classical characteristics in this case might pass the same point). In particular, considering the Cauchy problem  
(\ref{6eq1}), (\ref{6eq2}), we note that its solution  should be understood in the generalized sense, where
both the Rankine-Hugoniot condition \cite{Guenther1988}
\begin{eqnarray}
[u] \frac{d \sigma}{d t} = [f],
\label{6eq5}
\end{eqnarray}
where $[\cdot]$ denotes a jump across $x= \sigma(t)$ (i.e. $[u]=u_{\rm right} - u_{\rm left}$),
and the entropy condition
\begin{eqnarray}
a(u_{\rm right}) < \frac{d \sigma}{d t} < a(u_{\rm left}),
\label{6eq6}
\end{eqnarray}
meaning that the entropy increases as material crosses discontinuity,
 should be satisfied. 
Under these circumstances the development of constructive procedures for efficient numerical schemes for the solution of (\ref{6eq1}), 
 (\ref{6eq2}) becomes a very important task in theory and practice of conservation laws, as well as in a number of related fields. The stability of such schemes is at the heart of the success in achieving this task. Although viscosity solutions could provide some insight into these problems, they would not provide details on the solution stability.  Hyperbolic partial differential equation (PDE)  models are models for information propagations while probabilistic approaches imply diffusion. However, since such models need to be solved numerically, we will demonstrate that the stability conditions, derived in our approach in a straightforward manner, coincide with the stability conditions typical for numerical approximations of hyperbolic PDEs.


\subsection{Neural networks in approximating conservation law models}

There is an intrinsic analogy in the information flow pattern for neural networks used in applications (e.g., Fig. \ref{fig1}) and information flows represented by stencils of  numerical approximations of conservation laws (e.g., Fig. \ref{fig2}). Consider first some basic classical schemes for numerical approximations of  (\ref{6eq1}). We start from the forward centered Euler  scheme 
 \begin{eqnarray}
u_i^{j+1} = u_i^j - \frac{\lambda}{2}( u_{i+1}^j - u_{i-1}^j ). 
\label{6eq7}
\end{eqnarray}
 The flow of information for these schemes is given in Fig. \ref{fig2} (a) and it can be seen that in the terminology of neural networks the stencil for this approximation (as well as for an improved approximation $u_i^j \rightarrow \frac{1}{2} ( u_{i+1}^j + u_{i-1}^j )$) does not contain any hidden layers.
The situation becomes more involved for the  Leap-Frog scheme (see Fig. \ref{fig2} (b)), where  several layers of network architecture are coupled
\begin{eqnarray}
u_i^{j+1} = u_i^{j-1} - \frac{\tau}{h} (F_{i+1}^j - F_{i-1}^j).
\label{6eq8}
\end{eqnarray}
This coupling is sequential in nature, and hence our consideration is pertinent to  such networks architectures where the layers are {\em sequentially} linked.  From a numerical point of view such architectures can be represented by the predictor-corrector-type schemes. One of the most effective practical tool in numerical approximation of conservation laws that reflects this architecture is 
  the Lax-Wendroff scheme
\begin{eqnarray}
u_i^{j+1} = u_i^j - \lambda( F_{i+1/2}^{j+1/2} -  F_{i-1/2}^{j+1/2}), \quad 
\mbox{where} \quad F_{i+1/2}^{j+1/2} = \frac{1}{2} a u_{i+1}^j +  \frac{\lambda}{2} a^2 (u_{i+1}^j - u_i^j).
\label{6eq9}
\end{eqnarray}
As it is obvious from Fig. \ref{fig2} (c), this scheme includes ``hidden'' layer $j+1/2$ where one has to determine fluxes before moving to the next network layer. In what follows we will show that this scheme and many other effective schemes for numerical approximations of conservation laws can be 
 derived from the general probabilistic approach based on the association of the conservation law models with  Markov chain training processes for some neural network architectures in a way similar to that  described in Section 5. Moreover, the interpretation/association of the stencils of these schemes with neural network architectures will allow us to deal with the stability conditions in an efficient manner by using the consistency conditions of Markov chains with the original process.

{
\begin{center}
\begin{figure}
\centerline
{\epsfig{file=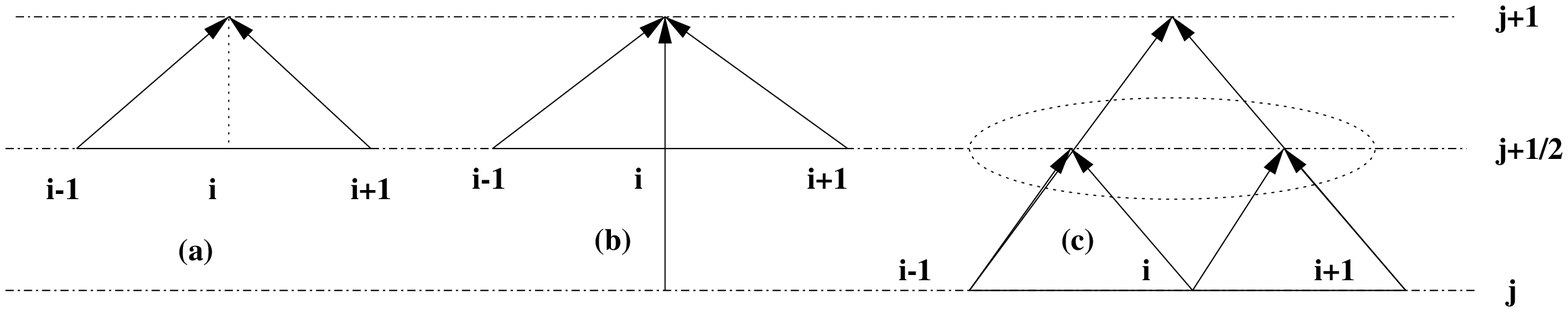, height=4cm, width=14cm}} 
\caption{Network architectures and information flow chart for numerical approximations 
 of conservation laws:  (a) forward centered Euler, (b) leap-frog, (c) Lax-Wendroff.}
\label{fig2}
\end{figure}
\end{center}
}

In the spirit of Section 5, the architecture of neural networks is associated with the cone of macroscopic events.  From the numerical point of view this is equivalent to the definition of  the following points
\begin{eqnarray}
& &
[(x_i, t_0^{\tau_{-1}}), i=0,...,2n] \rightarrow [(x_i, t_1^{\tau_{0}}), 
i=1,...,2n-1] \rightarrow  [ (x_i, t_2^{\tau_{1}}), i=2,...,2n-2]
\rightarrow ... \rightarrow 
\nonumber
\\[10pt]  
& & 
  [ (x_i, t_{n-1}^{\tau_{n-2}}), i=n-1,n,n+1] 
\rightarrow
[(x_i, t_n^{\tau_{n-1}}), i=n]
\label{6eq10}
\end{eqnarray}
 in the cone of macroscopic events, where the approximation of the evolution of dynamic systems described by the conservation law takes place. 
An important point to emphasize is that in this general framework the stability conditions for the associated schemes can be  obtained  in the explicit form. 
 Consider, for example, scheme (\ref{5eq18}), (\ref{5eq19}) for the solution of homogeneous problem (\ref{5eq4}),  (\ref{5eq5}). First note that if the following condition 
\begin{eqnarray}
v^+ (1 - \gamma_1 - 3 \gamma_2) + v^{-} (1 + \gamma_4 + 3 \gamma_3) = [v^{-}(1+ \gamma_3 - \gamma_4) - v^+ (1 + \gamma_1 - \gamma_2)]^2
\label{6eq11}
\end{eqnarray}
is met, then    the Courant-Friederichs-Lewy-type (CFL-type) stability condition 
(\ref{5eq17}) is satisfied \cite{Melnik1998}. Since either $v^-$ or $v^+$ is zero while the other is non-zero, we consider  2 cases. Our aim in this example is to derive explicit expressions for the flux limiters. In its turn, as follows from our discussion in Section 5, this will define stability conditions of the associated scheme and will  complete the definition of the transition probabilities of the associated Markov chain.

\noindent
{\bf 1.}
If $v^-=0$ and $v^+ \neq 0$ then from (\ref{6eq11}) we have
\begin{eqnarray}
1- \gamma_1 -3 \gamma_2 = (v^+)^2 (1 + \gamma_1 - \gamma_2)^2.
\label{6eq12}
\end{eqnarray}
This leads to a quadratic equation with respect to $\gamma_1$
\begin{eqnarray}
v^+ \gamma_1^2 +[2 v^+ (1 - \gamma_2)+1] \gamma_1 + [v^+(1- \gamma_2)^2 +3 \gamma_2 -1] =0.
\label{6eq13}
\end{eqnarray}
In a particular case where $\gamma_2=0$ the solution to  (\ref{6eq13}) can be easily found
\begin{eqnarray}
\gamma_{1}^{1,2} = \frac{-(2 v^+ +1) \pm \sqrt{8 v^+ +1}}{ 2v^+}.
\label{6eq14}
\end{eqnarray}
Since $\gamma_1$ should be non-positive (see (\ref{5eq14})) we take the sign ``minus''
in (\ref{6eq14}), leading to 
\begin{eqnarray}
\gamma_1 = -1 - \frac{\sqrt{8 v^+ +1} +1 }{2v^+}.
\label{6eq15}
\end{eqnarray} 
A similar reasoning leads to the expression for $\gamma_1$ in the general case of (\ref{6eq13})
\begin{eqnarray}
\gamma_1^{1,2} =  \frac{-[2 v^+(1- \gamma_2) +1] \pm \sqrt{8 v^+ +1 -16 v^+ \gamma_2} }{2 v^+}, 
\label{6eq16}
\end{eqnarray}
and finally to
\begin{eqnarray}
\gamma_1 = -1 + \gamma_2 - \frac{ \sqrt{8 v^+ +1 -16 v^+ \gamma_2} +1}{2 v^+}.
\label{6eq17}
\end{eqnarray}

\noindent
{\bf 2.} Along the same vein we can obtain the expression for $\gamma_4$ in the case where $v^+=0$ and $v^- \neq 0$ as a solution to the equation
\begin{eqnarray}
1+ \gamma_4 +3 \gamma_3 = v^- (1+ \gamma_3 - \gamma_4)^2 
\label{6eq18}
\end{eqnarray}
which can be written in a form easily amenable to its solution with respect to $\gamma_4$
\begin{eqnarray}
 v^- \gamma_4^2 -[2 v^-(1+\gamma_3)+1] \gamma_4 + [v^-(1+ \gamma_3)^2 -1 -3 \gamma_3] =0. 
\label{6eq19}
\end{eqnarray}
The solution to the last equation is determined as 
\begin{eqnarray}
\gamma_4 = 1 + \gamma_3 + \frac{\sqrt{8 v^- +1 + 16 v^- \gamma_3} +1 }{2 v^-}
\label{6eq20}
\end{eqnarray}
(the sign ``plus'' was taken due to the non-negativeness of $\gamma_4$, 
 see (\ref{5eq14})).
In the case $\gamma_3=0$ (\ref{6eq20}) is reduced to the expression obtained in \cite{Melnik1998}.


\begin{center}
\begin{figure}
\centerline
{\epsfig{file=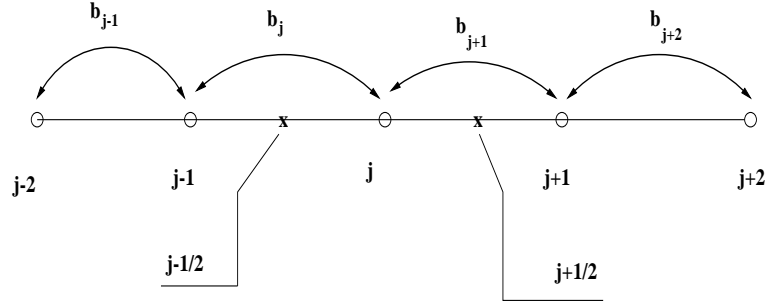, height=4cm, width=10cm}} 
\caption{Flux limiters in the probabilistic approach for conservation laws}
\label{fig3}
\end{figure}
\end{center}

This probabilistic approach based on the association of conservation law approximations and the neural network training process is amenable to the treatment of dissipative systems  by using the framework of perturbed generalized dynamic systems developed in \cite{Melnik1998} (see also references therein). However, to make  our basic concepts  transparent to the reader, we concentrate in this paper on dynamic systems without dissipation.
Since the representation (\ref{4eq1}) of dynamic system  (and its consequences) from a numerical analysis point of view is an explicit scheme, we recall \cite{Quarteroni2000,Lapidus1999} that a family of  explicit schemes  for the solution of (\ref{6eq1}), (\ref{6eq2}) can be written as
\begin{eqnarray}
u_j^{n+1}=u_j^n - \lambda (h_{j+1/2}^n - h_{j-1/2}^n),
\label{6eq21}
\end{eqnarray}
where $h_{j+1/2}^n=h(u_j^n, u_{j+1}^n)$ is the numerical flux that can model hidden layers in the network architecture associated with the given scheme.

We propose the following general approximation of the flux at $x_j + 0.5 \tau$ and $x_j - 0.5 \tau$, respectively
\begin{eqnarray}
&&
h_{j+1/2} = \sum_{k=j-1}^{j+2} b_k( \tau, h, v_k) u_k= 
b_{j-1} u_{j-1} +b_j u_j + b_{j+1} u_{j+1} + \tilde{b}_{j+2} u_{j+2}, 
\label{6eq22}
\\[10pt]  
&&
h_{j-1/2} = \sum_{k=j-1}^{j} b_k( \tau, h, v_k) u_{k-1}=
\tilde{b}_{j-1} u_{j-2} + b_j u_{j-1} + b_{j+1} u_j + b_{j+2} u_{j+1},
\label{6eq23}
\end{eqnarray}
where all coefficients in (\ref{6eq22}) and (\ref{6eq23})  are velocity-dependent approximations of the flux limiters, for example, $b_{j-1} \equiv b_{j-1}(\tau, h, v_{j-1})$, $\tilde{b}_{j-1} \equiv 
\tilde{b}_{j-1}(\tau, h, v_{j-2})$, etc.
Then, the general family of the explicit schemes can be written as follows
\begin{eqnarray}
u_j^{n+1} &=& u_j^n - \lambda [- \tilde{b}_{j-1} u_{j-2} + (b_{j-1}- b_j)u_{j-1}
+(b_j - b_{j+1}) u_j +(b_{j+1} - b_{j+2}) u_{j+1} +\tilde{b}_{j+2} u_{j+2}]  
\nonumber
\\[10pt]
&=&
u_j^n - \lambda \left[ - \tilde{b}_{j-1} u_{j-2} - \sum_{k=j-1}^{j+1} (b_k - b_{k+1}) u_k + \tilde{b}_{j+2} u_{j+2} \right].
\label{6eq24}
\end{eqnarray}
Note that this  technique is applicable to the case where 
$
b_m \equiv b_m ( \tau, h, \psi(v_{m-1}, v_m))$, e.g.  $\psi(v_{m-1}, v_m)= (v_{m-1}+ v_m)/2.$


\subsection{Special Cases and Examples}

Now, we demonstrate our  methodology on a number of examples and  show that most efficient schemes for numerical approximations of conservation laws become just special cases of our general network-based approach. 


\noindent
{\bf Example 1.} 
First of all, it is easy to confirm that  by choosing the flux limiters in the forms
\begin{eqnarray}
b_{j-1}=0, \quad b_{j-1}- b_j = - \frac{1}{2}a, \quad b_j - b_{j+1}=0, \quad  b_{j+1} -b_{j+2}=\frac{1}{2}a, \quad b_{j+2}=0,
\label{6eq25}
\end{eqnarray}
we obtain the forward centered Euler scheme for which the numerical flux is defined as
\begin{eqnarray}
h_{j+1/2}= \frac{1}{2} a (u_{j+1} + u_j).
\label{6eq26}
\end{eqnarray}


\noindent
{\bf Example 2.} 
If we take now
\begin{eqnarray}
\left [
\begin{array}{l}
b_{j-1}=0, \quad - \lambda(b_{j-1} - b_j) = \frac{1}{2} + \frac{\lambda}{2} a,
\quad 1- \lambda (b_j - b_{j+1})=0, 
\\[10pt]
-\lambda(b_{j+1}- b_{j+2}) = \frac{1}{2} - \frac{\lambda}{2} a, \quad 
b_{j+2}=0,
\end{array}
\right.
\label{6eq27}
\end{eqnarray}
we arrive at the following result   
\begin{eqnarray}
b_{j+1}= \frac{1}{2} a - \frac{1}{2 \lambda}, \quad b_j = \frac{1}{2} a - \frac{1}{2 \lambda}.
\label{6eq28}
\end{eqnarray}
Therefore, we confirm that under the above choice of the flux limiter coefficients, we obtain the Lax-Friedrichs scheme
\begin{eqnarray}
u_j^{n+1} = \frac{1}{2} (u_{j+1} + u_{j-1}) - \frac{\lambda a}{2} (u_{j+1} - 
u_{j-1}),
\label{6eq29}
\end{eqnarray}
which can be written in the form (\ref{6eq21}) with
\begin{eqnarray}
h_{j+1/2} = \frac{1}{2} \left[ a(u_{j+1} + u_j) - \frac{1}{\lambda} (u_{j+1} - u_j)  \right].
\label{6eq30}
\end{eqnarray}


\noindent
{\bf Example 3.} 
By choosing the flux limiters in the general architecture (\ref{6eq21}) -- (\ref{6eq23}) as  
\begin{eqnarray}
b_{j-1}=b_{j+2}=0, \quad b_{j-1} - b_j = - \frac{a}{2} - \frac{\lambda}{2}|a|,
 \quad b_j - b_{j+1}= |a|
\label{6eq31}
\end{eqnarray}
we also confirm that 
\begin{eqnarray}
b_j = \frac{1}{2}a + \frac{|a|}{2}, \quad b_{j+1}= \frac{1}{2}a - \frac{|a|}{2}.
\label{6eq32}
\end{eqnarray}
It can be seen that this leads to  the (upwind) Euler uncentered scheme, representable in the general form  (\ref{6eq21}) with 
\begin{eqnarray}
h_{j+1/2} = \frac{1}{2} \left [ a(u_{j+1} + u_j) - |a|(u_{} - u_j) \right].
\label{6eq33}
\end{eqnarray}


\noindent
{\bf Example 4.} 
Taking 
\begin{eqnarray}
b_{j-1}=0, \quad b_j = \frac{1}{2}a + \frac{\lambda a^2}{2}, \quad b_{j+1}= \frac{1}{2}a -
 \frac{\lambda a^2}{2}, \quad b_{j+2} =0
\label{6eq34}
\end{eqnarray}
in the general formulation (\ref{6eq21}) -- (\ref{6eq23}) 
leads to the one of the most popular schemes for the numerical approximations of conservation law (\ref{6eq3}), the  Lax-Wendroff scheme, discussed briefly earlier in this section,  where
\begin{eqnarray}
h_{j+1/2} = \frac{1}{2}  [ a(u_{j+1} + u_j) - \lambda a^2 (u_{j+1} - u_j  ].
\label{6eq35}
\end{eqnarray}

\vspace*{0.5cm}

\noindent
{\bf Example 5.} 
Finally, we note that the family of schemes proposed in \cite{Melnik1998} and briefly reviewed in Section 5 is also  a subset of the general representation given by (\ref{6eq21}) -- (\ref{6eq23}). Indeed, if we take 
\begin{eqnarray}
\left [
\begin{array}{l}
b_{j-1}= v, \quad b_{j-1} - b_j = - v^{+} (1 + \gamma_2) - v^{-} \gamma_4, \quad 
b_j - b_{j+1} = |v| + v^{-} \gamma_4 - v^{+} \gamma_1, 
\noindent
\\
b_{j+1} - b_{j+2} = - v^{-} (1 - \gamma_3) + v^{+} \gamma_1, \quad b_{j+2} = v^{-} \gamma_3,
\end{array}
\right.
\label{6eq36}
\end{eqnarray}
(as before, $\gamma_j$ are flux limiters of the scheme determined from the stability and consistency conditions), we can determine flux coefficient $b_{j+1}$ in two different ways. First,   ``moving from the left'' (see Fig. \ref{fig3}) it is easy to obtain the following chain of relationships:
\begin{eqnarray}
b_j =  v^{+}(1+ \gamma_2) + v^{-} \gamma_4, \quad 
b_{j+1} = b_j - |v| - v^{-} \gamma_4 + v^{+} \gamma_1 = v^{+}(1+2 \gamma_2 + \gamma_1) - |v|.
\label{6eq37}
\end{eqnarray}
On the other hand, ``moving from the right'' (see Fig. \ref{fig3}) we  get that 
\begin{eqnarray}
b_{j+1} = v^{-} \gamma_3 - v^{-} (1 - \gamma_3) + v^{+} \gamma_1 = v^{-} (2 \gamma_3 -1) + v^{+} \gamma_1.
\label{6eq38}
\end{eqnarray}
Therefore, by equating two expressions for $b_{j+1}$ we arrive at the conclusion that 
\begin{eqnarray}
v^{+} (1 + 2 \gamma_2) + v^{-} (1 - 2 \gamma_3) = |v|,
\label{6eq39}
\end{eqnarray}
which can be satisfied by setting $\gamma_2 = \gamma_3 = 0$. Having these flux limiters, $\gamma_1$ and $\gamma_4$ can be controlled by satisfying the stability conditions \cite{Melnik1998}.

\section{Stability and Consistency Requirements for Network-Based Approximations of Conservation Laws}

In the previous section we derived numerical approximations for conservation laws using the general network-based methodology. As we have discussed in  Section 4 this methodology originates from the probabilistic foundations of the dynamics. Indeed, having approximate initial data (e.g., as a result of measurements) we attempt to extrapolate this data further in time. This procedure is a subject of a probabilistic error. Therefore, it is important to discuss further stability and consistency requirements for the schemes obtained earlier in this paper, in particular, for the general family of approximations (\ref{6eq21}) -- 
(\ref{6eq23}).

The probabilistic approach to conservation laws allows us to explain  a number of important  phenomena related to numerical approximations of these models in a straightforward way. First, it is easy to check that the sum of all coefficients near the unknown function in every scheme discussed in the previous section is 1. However, this fact alone might not be sufficient to interpret those coefficients as transition probabilities. Indeed, the nonnegativity requirement plays a key role in the possibility of such an interpretation. From a numerical point of view this requirement leads to the stability conditions of the scheme. If we take the forward Euler/centered scheme this requirement leads to ${\lambda a}/{2} \leq 0$, which can be satisfied only if $a<0$ ($\lambda= \tau/h$). However, since we use here a forward scheme with a stencil depicted in Fig. \ref{fig2}, this should not come as a surprise, because for $a>0$ the explicit scheme on this stencil is known to be unstable.


It is easy to check that coefficients in all schemes considered in Sections 5 and 6 will be nonnegative subject to the Courant-Friedrichs-Lewy-type stability conditions $|\lambda a |<1$ (as one would expect for the models based on hyperbolic PDEs \cite{Quarteroni2000}) and therefore, all these schemes can be interpreted in a probabilistic sense.  Naturally, however, that satisfying those stability conditions is subject to  the appropriate choice of the flux limiters. For example, for the scheme (\ref{5eq18}), (\ref{5eq19}) discussed in Section 5, such fluxes ($\gamma_1$ and $\gamma_4$) should be chosen as  to  satisfy the following conditions
\begin{eqnarray}
\frac{\tau}{h}(|v|+v^{-} \gamma_4 - v^{+} \gamma_1  ) \leq 1, \quad \frac{1}{2} v^{+} + v^{-} \gamma_4 \geq 0, \quad  \frac{1}{2} v^{-} + v^{+} \gamma_1 \leq 0.
\label{7eq1}
\end{eqnarray}
Moreover, one has to satisfy the consistency conditions, which in the case of this scheme has the form
\begin{eqnarray}
\tau \left[ v^{-} (3/2 - \gamma_4) + v^{+}(3/2 + \gamma_1) - v  \right] =  {\it  o}( \tau + h),
\label{7eq2}
\end{eqnarray}
where, as before, ${\it o}$ denotes the Landau symbol.

\begin{center}
\begin{figure}
\centerline
{\epsfig{file=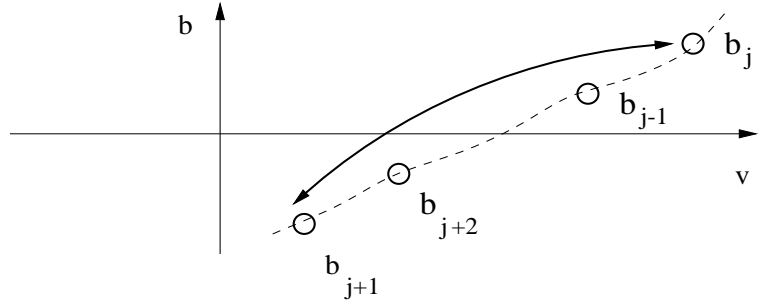, height=4cm, width=10cm}} 
\caption{A schematic representation of flux limiters as functions of velocity}
\label{fig4}
\end{figure}
\end{center}

\begin{center}
\begin{figure}
\centerline
{\epsfig{file=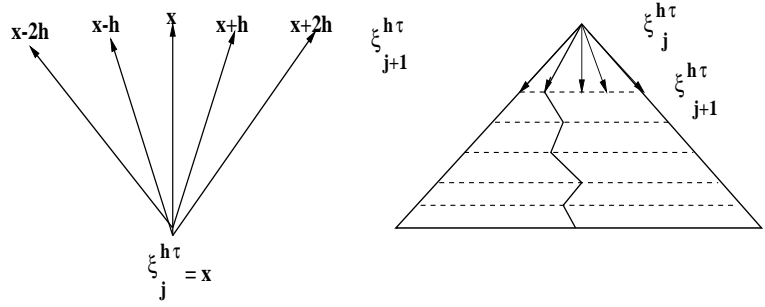, height=4cm, width=10cm}} 
\caption{Markov chain as a tool to combine feedforward and feedbackwards neural networks}
\label{fig5}
\end{figure}
\end{center}

In our general case of the family (\ref{6eq21}) -- (\ref{6eq23}) the nonnegativity of coefficients require
\begin{eqnarray}
\left [
\begin{array}{l}
\lambda b_{j-1} \geq 0, \quad \lambda (b_j - b_{j-1}) \geq 0, \quad  1 - \lambda(b_{j+1} - b_j) \geq 0, 
\\
\lambda(b_{j+2} - b_{j+1}) \geq 0, \quad - \lambda b_{j+2} \geq 0.
\end{array}
\right.
\label{7eq3}
\end{eqnarray}
This results in
\begin{eqnarray}
\lambda(b_{j+1} - b_j) \leq 1, \quad \mbox{where} \quad b_{j+1} \leq b_{j+2} \leq 0 \leq b_{j-1} \leq b_j
\label{7eq4}
\end{eqnarray}
(see Fig. \ref{fig4} for interpretation).
Consistency conditions for the general scheme (\ref{6eq21}) -- (\ref{6eq23}) can also be obtained by using the methodology developed in  \cite{Melnik1998}. Indeed, we assume that each jump of the associated Markov chain is allowed within a region (e.g., rectangular) with characteristic lengths $\tau$ and $h$ such that $\tau < h$ (if necessary such a region can be easily refined in a way similar, for example, to the finite element methodology). If the Markov chain is in state $x$ we associate this state with the position of the Markov chain, i.e. assume that $\xi_j^{h \tau} =x$. At the next moment of time $\tau \rightarrow \tau + \Delta \tau$ the Markov chain is assumed to be in one of the following states: $x-2h$, $x-h$, $x$, $x+h$, or $x+2h$ (see Fig. \ref{fig5}, left). This can be extended to the case of any arbitrary number of states in a straightforward manner. 
The table of transition probabilities from the old to a new state of the  Markov chain associated with our scheme (\ref{6eq21}) -- (\ref{6eq23})  can be constructed as follows. 

\vspace*{0.3cm}
\begin{tabular}{|l|l|}
     \hline 

     New state ($x_i$) & Probability of transition ($p_i$) \\
     \hline 

$x-h$ & $\displaystyle {\tau}/{h}(b_j - b_{j-1})$  \\
 \hline 

$ x+h $ & $\displaystyle {\tau}/{h}(b_{j+2} - b_{j+1})$ \\
     \hline 

$ x $  & $\displaystyle 1- {\tau}/{h}(b_{j+1} - b_j)$ \\
  \hline 

$x-2h$  & $\displaystyle {\tau}/{h}b_{j-1}$ \\
\hline  

$x+2h$  & $\displaystyle - {\tau}/{h}b_{j+2}$ \\
     \hline
     \end{tabular}
\vspace*{0.3cm}

We can now easily relate 
 this jump of the Markov chain to the transition to the next time layer in our numerical approximation of the conservation law. Therefore, we are in  a position to derive the  explicit form of the local consistency condition for the family of schemes (\ref{6eq21}) -- (\ref{6eq23}).
Since  
\begin{eqnarray}
E_j^{\tau h | (x, d^j)} = \sum_{i=1}^s x_i p_i,
\label{7eq5}
\end{eqnarray}
where $s$ is the number of states allowed in the next time layer (in our case $s=5$),
after some simplifications this condition can be given in the form
\begin{eqnarray}
 \tau \sum_{k=j-1}^{j+2} b_k = {\it o}(\tau + h). 
\label{7eq6}
\end{eqnarray}
The sign is chosen to be positive due to the direction of computations (see Fig. \ref{fig5}, right). 

For example,  consider the family of schemes discussed in Section 5 (a subclass of the general family (\ref{6eq21}) -- (\ref{6eq23})). Note that 
 in this case  
\begin{eqnarray}
\sum_{k=j-1}^{j+2} b_k = v^{+} [ 2(1 +2 \gamma_2) + \gamma_1 + \gamma_2] + v^{-} ( \gamma_3 + \gamma_4) - |v|. 
\label{7eq7}
\end{eqnarray}
Take, for example,  the limiters as $\gamma_1 = \gamma_2 = - 1/2$ and $\gamma_3 = \gamma_4 = 1/2$. This reduces the consistency condition to
\begin{eqnarray} 
-v^{+} + v^{-} - |v| = \sum_{k=j-1}^{j+2} b_k.
\label{7eq8}
\end{eqnarray}
Note also that the resulting scheme in this case has the following form
\begin{eqnarray} 
u_i^{j+1} = u_i^j +                                                                   \frac{\tau}{h} |v| \left [- \frac{3}{2} u_{i} +  \frac{1}{2} ( u_{i-1} + u_{i+1}) \right] +  \frac{\tau}{2 h} [v^{+}  u_{i-2} +  v^{-} u_{i+2}].
\label{7eq9}
\end{eqnarray}
It is easy to conclude that if  $v>0$ (and hence  $\displaystyle -2v = \sum_{k=j-1}^{j+2} b_k$) scheme (\ref{7eq9}) is reducible to 
\begin{eqnarray}
u_i^{j+1}= u_i^j + \frac{\tau}{2 h} v [-3 u_i^j + (u_{i+1} + u_{i-1}) + u_{i-2}],
\label{7eq10}
\end{eqnarray}
whereas if  $v<0$ (and hence $\displaystyle \sum_{k=j-1}^{j+2} b_k =0$) we have
\begin{eqnarray}
u_i^{j+1}=u_i^j -  \frac{\tau}{2 h} v [- 3 u_i + (u_{i+1} + u_{i-1})  
+ u_{i+2} ]. 
\label{7eq11}
\end{eqnarray}

The global consistency condition involves the velocity of the associated Markov chain  (see, e.g. (\ref{5eq14}) and further details in \cite{Melnik1998}). Recall that for the family of schemes constructed in \cite{Melnik1998}, the Markov chain velocity (that coincides with the velocity of the dynamic system/process when $n \rightarrow \infty$) is determined in terms of the flux limiters as follows
\begin{eqnarray}
v_{\rm MC} = v^{-} - v^{+} = -v.
\label{7eq12}
\end{eqnarray}
This result is independent of the sign of $v$. It  shows that the Markov chain evolves in the direction  opposite to the evolution of the dynamic system. This allows us to represent an explicit form of the global consistency condition for the family of schemes (\ref{6eq21}) -- (\ref{6eq23}) in  the form:
\begin{eqnarray}
\tau \left [ 3 h |v| - \tau v^2  \right ] = {\it o}(\tau +h).
\label{7eq13}
\end{eqnarray}

In conclusion, we note that, as follows from (\ref{7eq6}),  the Lax-Wendroff scheme's consistency condition takes the form
\begin{eqnarray}
\tau(b_j + b_{j+1}) = {\it o} (\tau +h) \quad \mbox{or} \quad \tau a  = {\it o} (\tau +h),  
\label{7eq14}
\end{eqnarray}
which leads to the CFL-type stability condition
\begin{eqnarray}
\frac{\tau a}{c (\tau +h)} \leq 1
\label{7eq15}
\end{eqnarray}
where  $\displaystyle c = \lim_{\tau \rightarrow 0, \; h \rightarrow 0} \frac{2 \tau v}{\tau + h}$ is a Landau constant participating in the  definition of probabilistic characteristics of conservation laws \cite{Melnik1998}. 
Consistency conditions for other schemes considered in Section 6 can be obtained in a similar manner.

\section{Conclusions}

In this paper, a general framework for the analysis of  a  connection between the training of artificial neural networks via the dynamics of Markov chains and the approximation of conservation law models has been proposed. 
This framework allows us to demonstrate an intrinsic link between microscopic and macroscopic models for evolution via the concept of perturbed generalized dynamic systems.
We have showed that mathematical models describing dynamics of network training can be treated effectively  by using 
 the concept of perturbed 
Markov chains associated with the original dynamic system. We have developed a general methodology allowing us to derive 
 computational models
for the dynamics of network training and to obtain  constructive algorithms, as well as  stability conditions for numerical approximations of conservation laws. We have demonstrated how this methodology can be applied to numerical approximations of conservation laws  viewed here as Markov chain network training procedures. Our main results have been  exemplified with several illustrative examples for which we have explicitly derived stability and consistency conditions.

\section*{Acknowledgments} I wish
 to thank my colleagues from Mathematics Departments at the University of South Australia  for fruitful discussions on the topics of this paper. Support of the Mads Clausen Foundation is also gratefully acknowledged.

\end{document}